# Title[1]: ALMA resolves the stellar birth explosion in distant quasar 3C298

P.D. Barthel[2], M.J.F. Versteeg, P. Podigachoski

*Kapteyn Astronomical Institute, University of Groningen, P.O. Box 800, 9700 AV Groningen, The Netherlands.*

M. Haas

*Astronomisches Institut, Ruhr-Universität Bochum, 44780 Bochum, Germany.*

B.J. Wilkes

*Harvard-Smithsonian Center for Astrophysics, 60 Garden Street, Cambridge, MA 02138, USA.*

C. de Breuck

*ESO, Karl-Schwarzschildstraße 2, 85748 Garching bei München, Germany.*

S.G. Djorgovski

*California Institute of Technology, MC 249-17, Pasadena, CA 91125, USA.*

---

**ABSTRACT**

Galaxies are believed to experience star formation and black hole driven nuclear activity symbiotically. The symbiosis may be more extreme in the distant universe, as far-infrared photometry with the Herschel Space Observatory has found many cases of ultraluminous cool dust emission in $z$>1 radio galaxies and quasars, which could have its origin in the central black hole activity, or in extreme starbursts. We here present strong evidence for an extreme circumnuclear starburst in the $z$=1.439 quasar 3C298. Our unparalleled 0.18 arcsecond resolution ALMA image at rest-frame 410µm wavelength shows that the ~40K dust in its host galaxy resides in an asymmetric circumnuclear structure. The morphology of this structure implies a starburst origin and a symbiotic physical relation with the AGN driven radio source. The symbiosis is likely to be a general property of distant massive galaxies.

# 1. Introduction

Given the well-known scaling relations between galaxies and their black holes (e.g., Ferrarese & Merritt 2000, McConnell & Ma 2013), galaxies are believed to experience star formation, i.e., conversion of gas into stars, and central black hole growth, i.e., AGN phenomena, symbiotically (Kormendy & Ho 2013). Indeed, evidence for such symbiosis has been building up: nearby QSOs prefer blue host galaxies (Trump et al. 2013), and several other nearby QSOs were observed by the Hubble Space Telescope to have circumnuclear emission line regions powered by star formation (Young et al. 2014). The symbiosis of black hole and global galaxy growth is even more intriguing because of the possible feedback effects: positive (AGN induced star formation) and/or negative (AGN quenching of star formation). Prime questions concern the physical AGN – star formation interplay: does it occur, where does it occur, and if so, when and how? To answer these questions, it is necessary to zoom in on the star formation in AGN hosts and conduct a spatial and/or kinematical study of the astrophysical interconnections.

We have embarked on an imaging study using the Atacama Large Millimeter/submillimeter Array, ALMA, in the Atacama desert of Northern Chile to zoom in on galaxy growth observed in the hosts of five compact, subgalactic-sized, high redshift 3C radio sources. This project is a follow-up of our Spitzer, Chandra, and Herschel studies addressing obscured AGN and star formation in the ultra-massive hosts of $z > 1$ 3C radio source (Haas et al. 2008, Barthel et al. 2012, Wilkes et al. 2013, Drouart et al. 2014, Podigachoski et al. 2015, 2016) – objects which have been and will continue to be landmarks in the study of active galaxies through cosmic time. Herschel photometry has shown that about one in three of these powerful AGN are radio-loud ULIRGs as inferred from their large cool dust masses, suggesting star formation rates, SFRs, of hundreds to over a thousand $M_\odot$/year.

Prime questions concern the nature of this cool dust, and its location: widespread in the AGN host galaxy, or localized and maybe somehow connected to the active nucleus? These massive dust reservoirs may be related to the "maximal starbursts" observed in distant, massive, non-active submillimeter galaxies, SMGs, which were found to be smaller than about 0.5arcsec (Tacconi et al. 2006). Our Herschel studies have also established the interesting trend that the cool dust luminosity is a function of the AGN age (Podigachoski et al. 2015), in the sense that old AGN – large double-lobed radio sources – are characterized by less dust emission than young ones with compact, subgalactic-sized radio sources. Within the starburst scenario, this would indicate positive feedback during the young AGN phase, negative feedback during its adult phase, or simply fading of the galaxy growth over time.

ALMA enables 0.15arcsec resolution imaging at 1mm wavelength, hence a spatial study of star formation related cool dust at the kpc-scale. This high resolution also permits optimal subtraction of the co-spatial non-thermal (synchrotron) emission from the radio source, using high-resolution radio images. Earlier ALMA imaging, at 0.7arcsec resolution, of mm-bright $z$~6 QSOs indicated the presence of marginally resolved circumnuclear cool dust, extending over 1-2kpc in their hosts (Wang et al. 2013). Compared to that study, our imaging reaches a factor of ~4 higher spatial resolution. This Letter presents the first sample source, 3C298, associated with a $z$=1.439 quasi-stellar object, QSO. This object, B1416+067, is a bright (V=16.79), red QSO with strong (REW 4.5A) associated CIV absorption (Anderson et al. 1987). The radio source 3C298 is a very luminous ($P_{178MHz} = 1.4 \times 10^{28}$ erg/sec) compact asymmetric triple radio source, classified as a Compact Steep-Spectrum (CSS) source. Its overall angular size is ~1.6arcsec (Pearson et al. 1985), which translates[3] to about 14kpc. High-resolution imaging (0.1arcsec) with MERLIN revealed that the central radio core breaks up into two components (Akujor et al. 1991): a core and jet knot, separated by 0.2arcsec. The radio source has strong spatial radio depolarization inhomogeneity (Mantovani et al. 2013). 3C298 is one of the strongest dust emitters in the 3C catalog, with significant Herschel detections at 250, 350 and 500μm (Podigachoski et al. 2015). The object was detected at 60μm by IRAS (Faint Source catalog), and also has single dish mm and submm detections (Willott et al. 2002, Haas et al. 2006). Decomposition of its infrared-mm SED indicates luminous, AGN-heated hot dust and luminous, presumably star formation heated cool dust (Podigachoski et al. 2015). The latter implies a well constrained, but model-dependent SFR of $940^{+30}_{-40}$ M$_\Theta$/year.

---

[3] We follow Bennett et al. (2014), using a $H_0 = 69.6$ km s$^{-1}$ Mpc$^{-1}$, $\Omega_\Lambda = 0.714$, and $\Omega_M = 0.286$ cosmology, which implies a scale of 8.6 kpc/arcsec

## 2. Observations and results

ALMA band 7 (1 mm) observations of 3C298 took place on August 15, 2016, with baselines of up to 1.6km, the phase calibrator J1410+0203, and an on-source integration time of 15minutes. Standard CASA pipeline calibration was employed at the ALMA European Regional Centre Node in Leiden, The Netherlands. Several steps of CASA imaging were carried out, including phase-only and phase+amplitude self-calibration. The resulting beam is 0.18 x 0.16 arcsec in p.a. -5°, and the final 1mm image, shown in Figure 1a, reaches a $1\sigma$ noise level of 37 μJy/beam. 3C298 is an asymmetric, slightly bent triple source, with a bright, marginally resolved, central component of 18.9±0.04 mJy[4]. We observe mm lobe radiation about 1arcsec to the East, and a Western lobe much closer in, at 0.4arcsec from the nucleus.

---

[4] Quoted 1-sigma errors represent the combination of image and component fitting noise; formally there is another 5% ALMA absolute flux density uncertainty

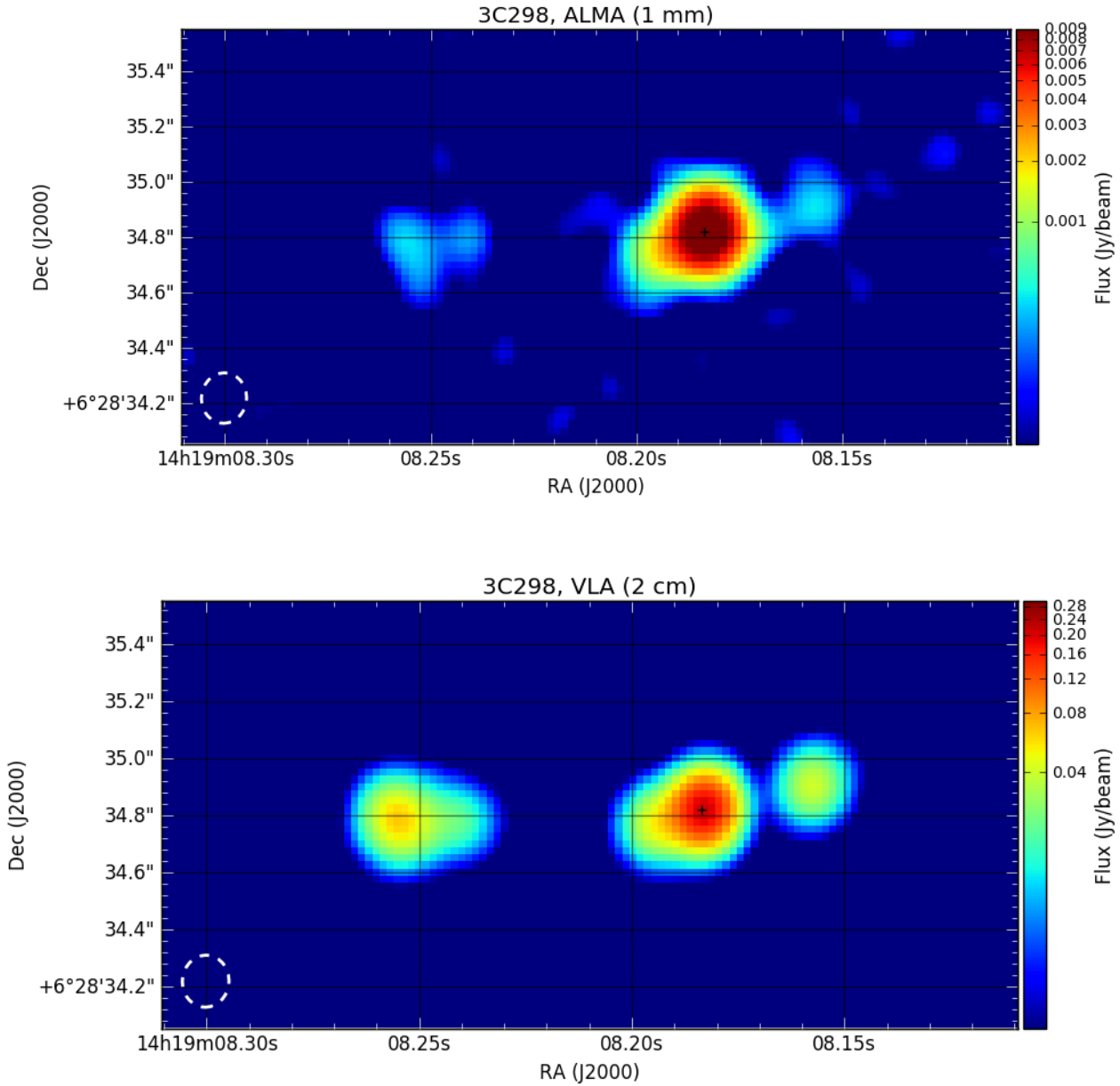

Figures 1a – 1b The ALMA 1mm image of 3C298 and its VLA 2cm image, at identical resolution of 0.18 x 0.16 arcsec in p.a. -5°, top and bottom, resp. The resolving beams are shown in the bottom-left corners; the peak flux densities are 0.0146 and 0.210 Jy/beam, for the 1mm and 2cm image, respectively. The ALMA image 1σ noise is 37 μJy/beam, the VLA image 1σ noise is 0.87 mJy/beam. The +-signs indicate the position of the optical QSO.

As alluded to above, the mm radiation from the nucleus of a radio-loud object consists of two parts: the thermal Rayleigh-Jeans tail of the cool (30-40K) host galaxy dust component and the synchrotron component of its radio source, the strength of which can be extrapolated from the shape of its radio spectrum. There is also a third component in the form of free-free radiation, but its magnitude is insignificant at 410µm rest-frame wavelength (Condon 1992). To establish the strength and morphology of the cool dust thermal emission in the 3C298 host, that is its Rayleigh-Jeans tail at 1mm observed wavelength, we proceeded as follows. We started by analysing VLA Archive data (project BREUGL) at U-band (2cm, 15GHz) of 3C298, taken in October 1983. These observations were carried out in the A-array, yielding an angular resolution of 0.17 x 0.13 arcsec in p.a. -54°, i.e., very similar to our ALMA imaging. Standard AIPS routines including a small final convolution were employed to yield the VLA 2cm image as shown in Figure 1b, having a noise of **0.87**mJy/beam and resolution identical to the ALMA 1mm image. Both images, 1a and 1b, were re-gridded to 0.025arcsec pixels. Their peak values (core and hot spots) coincide to within a pixel, indicating that the 1mm image is dominated by synchrotron emission. We applied small, milliarcsecond precision shifts to both images, aligning their unresolved peaks to the optical QSO position (using its Gaia position $14^h19^m08^s.1835$, $+06°28'34.819''$ (J2000)).

Visual inspection of the 1mm image suggests a slightly extended circumnuclear morphology underlying the central core-jet structure seen on the VLA 2cm image. This underlying structure could be due to thermal emission. We investigated its properties by subtracting the cm image from the mm image. First of all, we clipped the two images at their respective 3σ level. Besides accurate alignment, subtraction requires an estimate of the spectral index between the two data sets, accounting for possible spectral index gradients. Considering the outer lobes including the hot spots, we find 2cm to 1mm spectral indices of -1.6 and -1.5 for their integrated emission, E and W respectively. The central structure is expected to have a flatter spectral index. Ensuring milliarcsec peak alignment, we subtracted the 2cm image from the 1mm image using a range of spectral indices, reasoning as follows.

The ALMA peak is obviously not 100% thermal, as already mentioned. While this is unlikely (cf. Haas et al. 2006), let us suppose it is 0% thermal, or 100% synchrotron. This implies a (pure synchrotron) spectral index $\alpha_{2cm}^{1mm}$ = -0.9 ($S_\nu \propto \nu^\alpha$). Adopting that scaling factor for the subtraction of the 2cm emission from the 1mm emission yields the image, shown in Figure 2a, having zero central intensity and a marginally resolved clump of strength 0.70±0.07 mJy with a sharp edge, about 0.1arcsec West of the peak/QSO location. That structure is part of an extended plateau of size ~0.5arcsec (~4.5kpc), with a strong negative jet residual, caused by too much subtracted jet emission. The plateau has a secondary peak at the end of a curved tail-like extension, located south of the end of the synchrotron jet. Choice of a synchrotron spectral index $\alpha_{2cm}^{1mm}$ = -1.0 yields a more luminous (3.1±0.06 mJy) and larger, more "organic" thermal structure, now with nonzero flux at the QSO position, still part of the extended ~4.5kpc plateau with the tail-like extension. That thermal structure is shown in Figure 2b. The depression East of the peak is due to the subtraction of too much (steeper spectrum) jet emission. Image subtraction with spectral indices steeper than -1.0 yields pronounced residual emission at the QSO position, presumably non-thermal. The fact that the 3C298 2cm core has a steep synchrotron spectrum with spectral index of at least 0.9 indicates unresolved jet emission (indeed observed with VLBI – Fanti et al. 2002), and furthermore that any free-free radiation contribution is negligible.

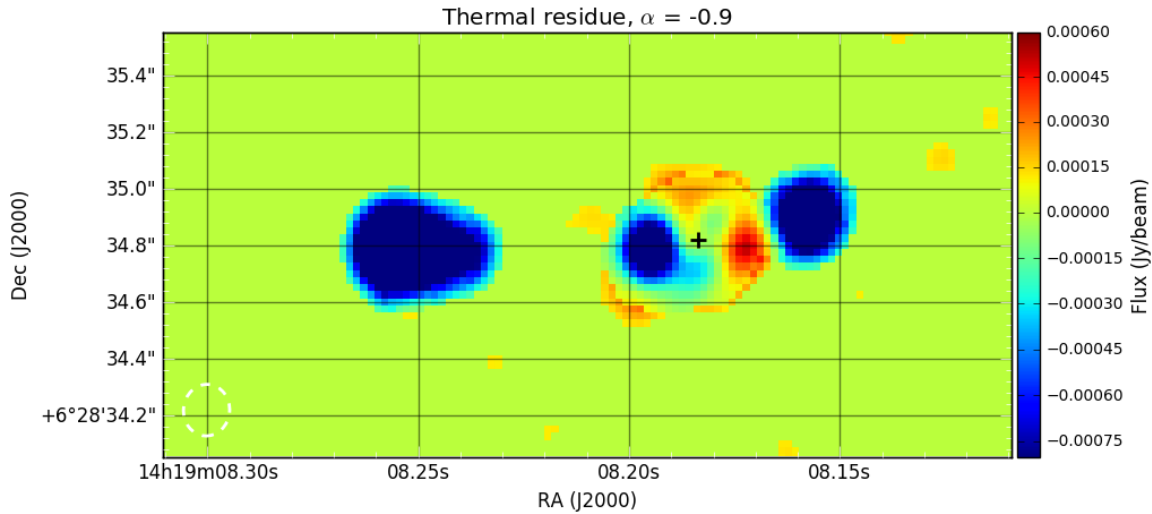

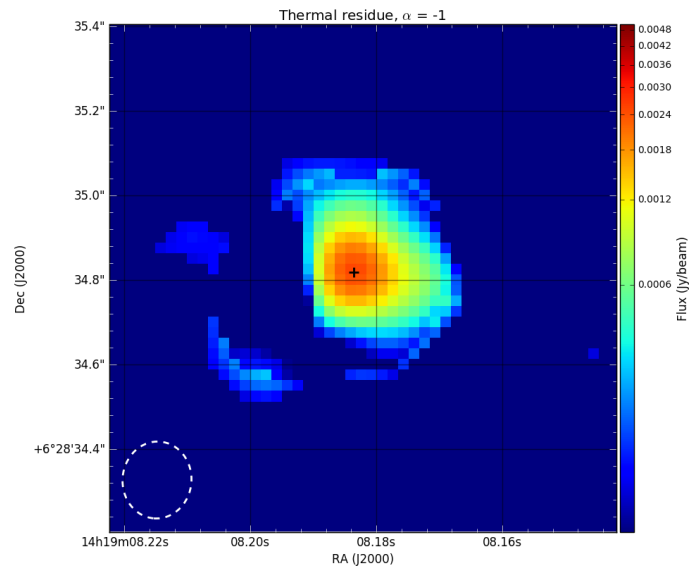

**Figures 2a – 2b.** Two representations of the (1mm minus 2cm) residual structure in 3C298 – see text. The top image, having a noise of 71 µJy/beam, results from adopting a spectral index -0.9. The blue features are strongly negative and result from subtraction of too much synchrotron emission. The bottom image, having a noise of 57 µJy/beam, shows the zoomed-in thermal residual structure, i.e., the circumnuclear cool dust emission in the 3C298 host galaxy, obtained adopting a spectral index -1.0. The +-signs indicate the position of the non-thermal peak, coinciding with the optical QSO position.

Hence, although its precise morphology remains to be determined, 3C298 has asymmetrically extended, circumnuclear thermal 1mm emission, of strength at least 0.7mJy but more likely ~3.0mJy. The latter implies that ~16% of the core 1mm flux density of 18.9mJy resides in the thermal structure, while the remaining ~84% accounts for ~15.8 mJy nuclear (core+jet) synchrotron radiation. Of course there may be additional thermal emission under the jet. Our ALMA measured thermal core contribution is most likely larger than the 2.1mJy earlier estimated using a simple $T = 50$K grey body fit to the 450μm and 850μm flux densities (Haas et al. 2006). This is not unexpected as the actual cool dust temperature appears to be lower: Podigachoski et al. (2015) find 37.9K, implying more thermal radiation at 1mm. In the meantime, our ALMA-determined thermal 1mm flux density does not change the value of the SED-inferred SFR of $940^{+30}_{-40}$ $M_\odot$/year (Podigachoski et al. 2015), which combine the FIR photometry – most importantly the Herschel SPIRE flux densities – with the earlier, less precise mm estimates.

## 3. Discussion

Figure 3 presents the cool dust emission in 3C298 with the contours of the high-resolution VLA 2cm image overlaid. As described above, the images were aligned using the location of the unresolved synchrotron peak. We conclude that the coeval starburst in the 3C298 host galaxy occurs in an extended kpc-sized region and note with interest that these dimensions are similar to those of the circumnuclear star-forming rings observed in nearby low luminosity AGN hosts (Barth et al. 1995, Colina et al. 1997), and the dust structures observed in very distant QSOs (Wang et al. 2013).

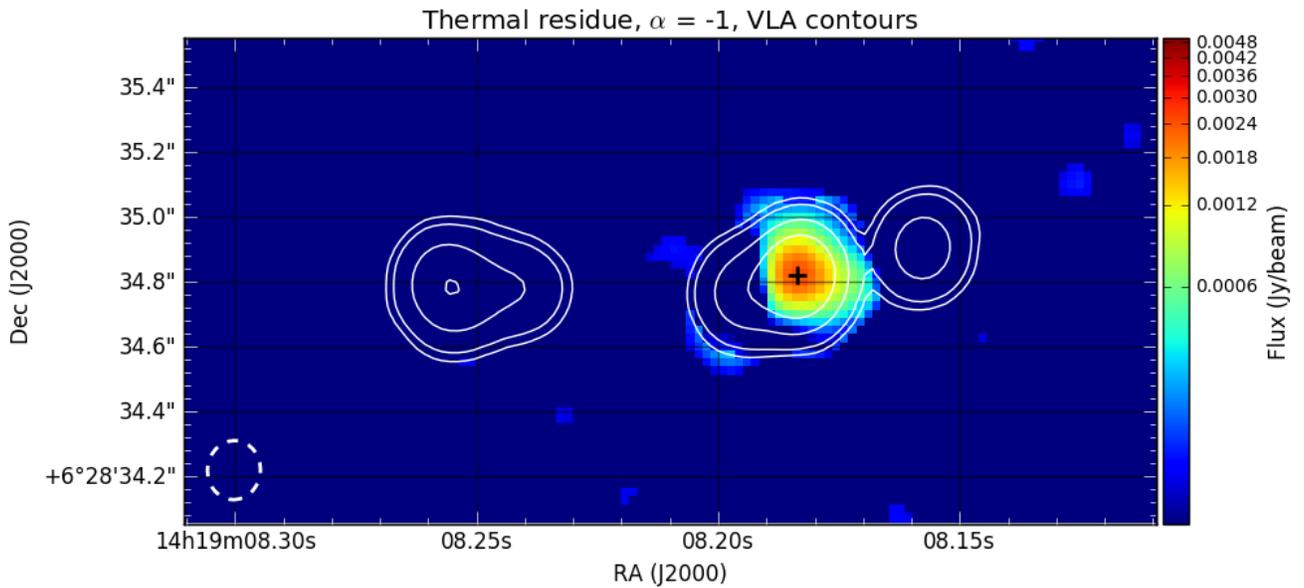

**Figure 3:** overlay of the asymmetric cool dust structure in the central regions of the 3C298 host galaxy, and the VLA 2cm image contours of its radio source. The contours are at 3, 6, 24, 72 x 0.87 mJy/beam.

Two possible occurrences of radio jet – starburst interaction in 3C298 are noteworthy. Firstly, we observe a clump of cool dust about 0.2arcsec towards the South-East, where a radio knot (visible on the 6cm MERLIN image (Akujor et al. 1991)) marks the deflection of the jet towards pure-East. This feature recalls jet-triggered star formation (Dey et al. 1997) but could also represent an ISM density gradient with associated star formation, causing jet deflection. As seen from Figure 3 (and on the MERLIN 6cm image), the 0.6arcsec triple radio source centred on the 3C298 nucleus is well aligned; its Eastern end, 0.2arcsec East-Southeast from the nucleus, marks the jet deflection location towards the Eastern lobe. Hubble Space Telescope rest-frame ultraviolet and visible observations have shown (Hilbert et al. 2016) that the host galaxy of 3C298 has a disturbed structure towards the E-SE. Secondly, the star formation plateau is asymmetrically extended towards the West. That is also the direction towards the closer lobe, which is, at ~3kpc distance, probably still within the host galaxy. This suggests an ISM density enhancement towards the West resulting in jet induced star formation or simply jet advancement frustration. On the other hand, we cannot rule out projection effects of a symmetric triple radio source, with the quasar radio axis observed at a small angle, its Eastern jet approaching, and its Western side observed at an earlier stage, hence closer to the core due to differential light travel time effects. In fact, Drouart et al. (2012) propose a jet inclination of 31° for the 3C298 radio source. Kinematic data at increased spatial resolution and starburst age dating are required to firmly establish the presence or absence of interaction and physical feedback. While 2D-spectroscopy of the 3C298 host galaxy at near-IR and mm-wavelengths was recently claimed to demonstrate galactic scale feedback driven by the AGN (Vayner et al. 2017), we argue that those observations can equally well be explained within the nuclear starburst (+ wind) scenario advocated here. In fact, the ionized gas winds reported from those observations (Vayner et al. 2017) show no clear connection to the radio jets, the reported CO-disk (their Fig. 7) coincides with and has comparable size and asymmetry to the circumnuclear dust structure which we find[5], and they together support the central starburst hypothesis. We also note that the cool dust emission being attributed to excessive kpc-scale host star formation supports the hypothesis of starburst-driven superwinds being the origin of associated heavy element absorption in QSO optical-uv spectra (Heckman et al. 1990, Barthel et al. 2017), which is a key feature of the present reddened QSO 3C298 (Anderson et al. 1987).

Finally, and most importantly, our identification of the cool dust emission as originating from a starburst gives confidence in that mechanism being the hitherto tacitly assumed (e.g., Drouart et al. 2014, Leipski et al. 2014, Ma & Yan 2015, Podigachoski et al. 2015, 2016, Pitchford et al. 2016, Westhues et al. 2016) source of the long-wavelength far-infrared emission. There is no doubt that early-epoch AGN hosts are very dynamic systems (Dey et al. 1997, Vayner et al. 2017, Nesvadba et al. 2017), and indeed our ALMA imaging has now observed a stellar birth explosion in the circumnuclear regions of the extremely powerful AGN 3C298. Higher resolution ALMA studies must be sought to reveal the morphological details of the starburst, while high resolution imaging with the forthcoming James Webb Space Telescope may reveal the newly formed star clusters.

---

[5] ALMA does not observe dust at 16kpc SE, where Vayner et al. (2017) find another CO clump

# 4. Conclusions

The suspected – on the basis of its SED – extreme starburst in the host galaxy of quasar 3C298 is indeed observed, and found to occupy its central regions. Its morphology implies a symbiotic physical relation with the AGN driven radio source, the precise nature of which remains to be determined. Our ALMA image proves that attributing the cool thermal emission in the 3C298 SED to starburst heating is correct. Given the frequent incidence of ultraluminous T~40K dust emission in high redshift AGN, the symbiotic occurrence of starbursts and black hole accretion is likely to be widespread.

The assistance of Drs. L. Guzman-Ramirez and Y. Contreras at Allegro, the European ALMA Regional Centre Node in Leiden, and of Groningen BSc D. Bartels is gratefully acknowledged. PDB acknowledges the hospitality of NRAO Socorro, where the manuscript was finalized. BJW acknowledges the support of NASA Contract NAS8-03060 (CXC). SGD was supported in part by the NSF grants AST-1413600 and AST-1518308, and by the Ajax Foundation. We acknowledge careful reading and insightful comments by the referee. This paper makes use of the following ALMA data: ADS/JAO.ALMA#2015.1.00754.S. ALMA is a partnership of ESO (representing its member states), NSF (USA) and NINS (Japan), together with NRC (Canada), NSC and AS/AA (Taiwan), and KASI (Republic of Korea), in cooperation with the Republic of Chile. The Joint ALMA Observatory is operated by ESO, AUI/NRAO and NAOJ.